\documentclass{article}
\usepackage{arxiv}
\usepackage[utf8]{inputenc}
\usepackage[T1]{fontenc}
\usepackage{hyperref}
\usepackage{url}
\usepackage{booktabs}
\usepackage{graphicx}
\graphicspath{{assets/}}
\usepackage{listings}
\usepackage{xcolor}
\usepackage{amsmath}
\usepackage{enumitem}
\usepackage{caption}
\usepackage{float}
\usepackage{placeins}

\lstdefinestyle{sql}{
  language=SQL,
  morekeywords={SELECT,FROM,JOIN,WHERE,ORDER,BY,DESC,LIMIT,AND,AS,IN,GROUP,MAX,ON,LENGTH},
  keywordstyle=\color{blue!70!black},
  stringstyle=\color{green!50!black},
  commentstyle=\color{gray!60},
}

\lstdefinestyle{python}{
  language=Python,
  morekeywords={embed,candidate_matrix,query_vec,embeddings,scores,direction,centroid,alpha},
  keywordstyle=\color{blue!70!black},
}

\newcommand{\flexvec}{\textsc{flexvec}}
\newcommand{\flex}{\textsc{flex}}

\title{\flexvec: SQL Vector Retrieval with Programmatic Embedding Modulation}

\author{
  Damian Delmas \\
  Independent Researcher, Vancouver, BC \\
  \texttt{damian@getflex.dev}
}

\begin{document}
\maketitle

\begin{abstract}
As AI agents become the primary consumers of retrieval APIs, there is an opportunity to expose more of the retrieval pipeline to the caller. \flexvec{} is a retrieval kernel that exposes the embedding matrix and score array as a programmable surface, allowing arithmetic operations on both before selection. We refer to composing operations on this surface at query time as Programmatic Embedding Modulation (PEM). This paper describes a set of such operations and integrates them into a SQL interface via a query materializer that facilitates composable query primitives. On a production corpus of 240,000 chunks, three composed modulations execute in 19 ms end-to-end on a desktop CPU without approximate indexing. At one million chunks, the same operations execute in 82 ms.
\end{abstract}

\section{Introduction}

Retrieval APIs are increasingly consumed by AI agents rather than humans. Most existing systems expose search as a fixed endpoint with its own query language, filter syntax, and result format. An agent can compose SQL directly when given a schema; SQL is already well-represented in training data, requires no new DSL, and supports filtering, joins, grouping, and error recovery natively.

These systems typically expose neither the embedding matrix nor the score array to the caller. \flexvec{} exposes both as a programmable surface on which arithmetic operations compose before selection, including suppression, diversity, trajectory, centroid expansion, and temporal decay. We refer to composing operations on this surface as Programmatic Embedding Modulation (PEM). A query materializer integrates these operators into SQL as composable query-time primitives, without modifying the database engine.

This design requires the embedding matrix in memory. The production corpus used in this paper is AI coding session history: 240,000 chunks (each chunk is one indexed unit: a message, tool call, or file snapshot) across 4,000 sessions with structured metadata including message type, project, timestamps, file paths, and tool names, stored in a single SQLite database.

\flexvec{} is evaluated here as the retrieval kernel within \flex{}, a local search and retrieval system where an AI agent connects via MCP (Model Context Protocol), discovers the schema at runtime, and writes read-only SQL against it. The system has been in production use since early February 2026, serving approximately 6,500 agent queries over six weeks.

This paper makes three contributions:

\begin{enumerate}[leftmargin=*]
\item \textbf{A query materializer architecture} that integrates external operators into SQL as composable query-time primitives, without modifying the database engine. Pseudo-functions are intercepted, executed outside the SQL engine, and materialized as temp tables before SQLite sees the query. This enables arbitrary SQL to determine which candidates enter scoring.

\item \textbf{Programmatic Embedding Modulation (PEM)} as the operator model within this architecture: the embedding matrix and score array become a programmable surface on which arithmetic operations compose before selection.

\item \textbf{A practical evaluation} demonstrating algebraic correctness of modulation formulas, composability, and millisecond-scale latency with composed modulations over corpora up to one million chunks using in-process numpy and SQLite.
\end{enumerate}

\section{Preliminaries}
\label{sec:prelim}

\subsection{Embeddings and Similarity}

An embedding model maps a text segment to a fixed-length vector. For L2-normalized vectors, cosine similarity reduces to a dot product; scoring $N$ candidates against one query is a single matrix multiply. Standard retrieval takes the top $K$ from the resulting scores. This baseline, brute-force cosine similarity over the candidate matrix, is the foundation on which the architecture and modulations build. Each chunk has a precomputed 128-dimension embedding (Nomic Embed v1.5~\cite{nomicembed2024}, Matryoshka truncation~\cite{matryoshka2022}).

\subsection{Filtered Retrieval}

Approximate nearest-neighbor indexes (HNSW, IVF) accelerate similarity search but introduce a tradeoff when combined with metadata filters~\cite{filteredann2025}. Pre-filtering requires per-predicate index construction. Post-filtering loses recall when the passing subset is small: the index wastes work on candidates that do not survive the filter. Existing solutions tend to handle specific filter types or selectivity ranges, but not both. These approaches target large-scale settings.

For corpora that fit in memory, brute-force matrix multiplication scores all candidates directly and can operate in under 100 ms. Without an index, the embedding matrix remains accessible as an array, and operations beyond top-$K$ similarity (subtraction, decay, centroid shifting) become straightforward at query time. PEM requires this direct access to the embedding matrix, which most ANN indexes do not expose.

\subsection{Operations on Embeddings}

The discovery that vector operations on word embeddings encode semantic relationships, such as \emph{king $-$ man $+$ woman $\approx$ queen}~\cite{mikolov2013}, established that embedding spaces support algebraic manipulation. Modern sentence embeddings inherit this property: adding a centroid shifts the query toward a cluster, blending two directions steers between topics. Other operations act on scores after the dot product: subtracting directional similarity suppresses a concept, weighting by timestamp favors recency.

\begin{itemize}[leftmargin=*]
\item \textbf{Query expansion (Rocchio).} Shifting the query toward the centroid of known-relevant documents~\cite{rocchio1971}. \flexvec{} uses positive-only centroid blending inspired by Rocchio, without the negative document term.

\item \textbf{Maximal Marginal Relevance (MMR).} Iterative selection that penalizes candidates similar to already-selected results, trading relevance for diversity~\cite{carbonell1998}.

\item \textbf{Contrastive query optimization.} DEO~\cite{deo2026} decomposes queries into positive and negative components and optimizes the query embedding at inference time, achieving negation-aware retrieval without training.
\end{itemize}

\section{Method}
\label{sec:method}

The goal is to let an AI agent search a corpus by writing SQL in response to user requests. A typical use for our production corpus involves questions about previous sessions, the lineage of a file within a codebase, or practical information from days or weeks prior.

The agent begins by viewing the schema (\S\ref{sec:schema}), then performs a broad semantic search and narrows based on what it finds: pre-filtering by project, suppressing dominant clusters, or switching to keyword search. For each query, it writes one statement that composes filtering, scoring, and post-processing, and the system executes it against a single SQLite database. The returned chunks remain in the session context, and the agent synthesizes a response.

This section describes \flexvec{}'s architecture as deployed within \flex{}, where the agent accesses it via an MCP server. The remainder covers how the agent discovers the schema (\S\ref{sec:schema}), how queries flow through three phases (\S\ref{sec:arch}), how modulations compose on scores (\S\ref{sec:modulations}), and the interface layers available to the agent (\S\ref{sec:interface}).

\subsection{Schema as Interface}
\label{sec:schema}

To guide the AI agent, instructions are split across two layers. Static MCP instructions ($\sim$2,200 tokens) are injected into the agent's system prompt, describing the retrieval pipeline, token grammar, and query methodology. At runtime, the agent calls \texttt{@orient}, a SQL preset that queries the live database for its own schema: tables, columns, available functions, and presets. Because the preset uses \texttt{pragma\_table\_info()} to discover view columns at runtime, schema changes propagate without updating the instructions. The MCP instructions are in Appendix~B; the \texttt{@orient} preset and output are in Appendices~C and~D.

\subsection{Architecture}
\label{sec:arch}

The agent discovers the schema (\S\ref{sec:schema}) and writes SQL. \flexvec{} bridges numpy into a SQL interface via a query materializer that turns a single statement into a three-phase pipeline. The host process loads embeddings into a numpy matrix at startup. The following example illustrates all three phases in a single statement, where an agent searches for system architecture while suppressing descriptive content that would otherwise dominate the results:

\vspace{2pt}
\noindent
\begin{minipage}{\linewidth}
\begin{lstlisting}[style=sql]
SELECT v.id, v.score, m.content
FROM vec_ops(
    'similar:how the system works architecture
     diverse
     suppress:website landing page design tagline
     suppress:documentation readme community post',
    'SELECT id FROM messages
     WHERE type = ''assistant'' AND length(content) > 300') v
JOIN messages m ON v.id = m.id
ORDER BY v.score DESC LIMIT 5
\end{lstlisting}
\end{minipage}

The inner subquery (\texttt{SELECT id FROM messages WHERE ...}) is Phase~1, pre-filtering to assistant messages longer than 300 characters. The \texttt{vec\_ops()} call is Phase~2, scoring candidates and applying suppression and diversity. The outer \texttt{SELECT ... JOIN ... ORDER BY} is Phase~3, composing the final result. This query is evaluated in detail in Section~\ref{sec:suppress}.

Each phase has a distinct role:

\textbf{Phase 1: SQL pre-filter.} A subquery returns the chunk IDs that enter scoring, scoped by project, time range, message type, or any combination the agent needs.

\textbf{Phase 2: numpy modulation.} Base cosine similarity via matmul, followed by optional modulations that reshape scores before the top $K$ are selected. This stage passes 500 scored candidates to Phase~3.

\textbf{Phase 3: SQL compose.} The outer query joins the scored candidates back to the database (content, metadata, any table the schema exposes) for grouping, reranking, or filtering.

\begin{figure}[H]
\centering
\includegraphics[width=\textwidth]{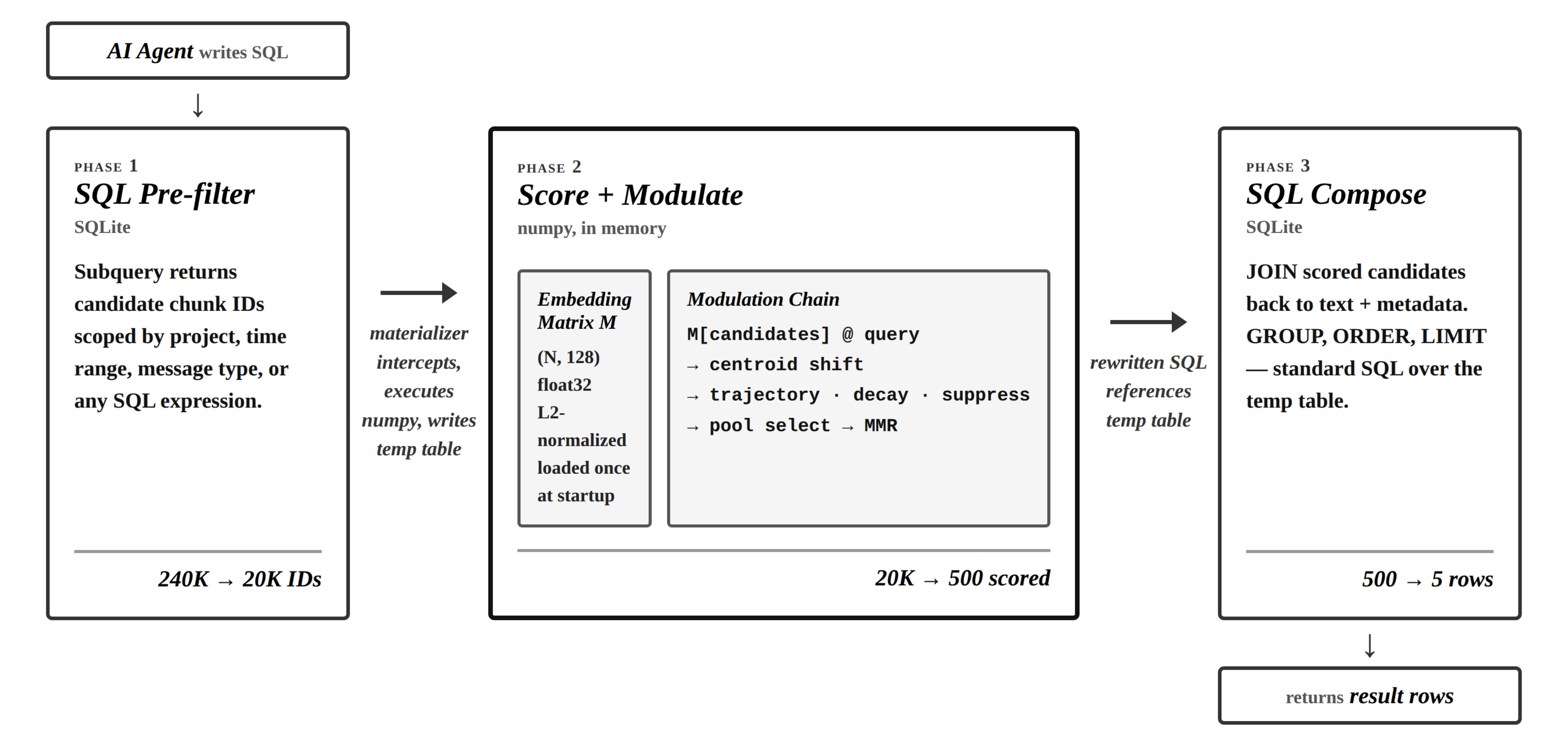}
\caption{An agent's SQL statement flows through three phases (pre-filter, score, compose) and returns ranked chunks.}
\label{fig:pipeline}
\end{figure}

Figure~\ref{fig:pipeline} illustrates these phases on the example query above, where Phase~1 pre-filters 240,000 chunks to approximately 20,000 candidates via SQL, Phase~2 scores those 20,000 embeddings and writes the top 500 to a temporary table, and Phase~3 joins those 500 candidates back, returning the top 5.

Pre-filtering reduces the candidate set before scoring, so modulations operate on fewer vectors. Because Phase~2 operates on a numpy array, the operator surface extends beyond scoring: structural operators such as per-query clustering and centrality can compute over the same candidates and expose results as additional columns for Phase~3.

The agent writes a single SQL statement. \texttt{vec\_ops()} and \texttt{keyword()} are not SQLite virtual tables or extensions; they are pseudo-functions recognized by a materializer that intercepts the query before execution. The agent's statement describes what to retrieve, and the materializer decides how to execute it: dispatching each pseudo-function to the appropriate engine (SQLite for pre-filter and composition, numpy for scoring and modulation, FTS5 for keyword search), writing results to a temporary table, and rewriting the SQL to reference it before handing it back to SQLite. Limitations of this approach are discussed in Section~\ref{sec:limitations}.

Adding a new query-time operator requires only implementing its execution over the candidate matrix and registering its token-level interface ($\sim$15 lines of Python). Materializers that introduce new execution contexts are larger in our implementation ($\sim$150 lines). \texttt{keyword()}, for example, was originally raw FTS5 SQL that leaked infrastructure to the agent and was converted to a materializer to absorb that complexity.

Phase ordering is enforced by the materializer rather than delegated to SQLite's planner, which enables the three-phase pipeline described above.

\subsection{Modulations}
\label{sec:modulations}

Standard vector retrieval scores candidates and selects the top $K$ in a single step. The score array --- the output of the matrix multiply --- is typically used only to select the top $K$ and is not exposed as a programmable object. PEM exposes that array as a programmable surface: between scoring and selection, operations reshape it. Topics are suppressed, recency is weighted, and the query shifts toward known-good examples.

Each modulation transforms scores. They compose in sequence: the output of one is the input to the next. Because each modulation transforms the scoring function, absolute scores are not comparable across different modulation configurations. The ranking within a single configuration is meaningful; the magnitude is not.

\begin{table}[H]
\centering
\small
\resizebox{\textwidth}{!}{%
\begin{tabular}{lllll}
\toprule
Token & Operation & Code & Cost & Ref \\
\midrule
\texttt{suppress:X} & Subtract similarity to suppression direction & \texttt{scores -= w * (M @ embed(X))} & 1 matmul & --- \\
\texttt{decay:N} & Reciprocal decay with $N$-day half-life & \texttt{scores *= 1/(1 + days/N)} & elementwise & --- \\
\texttt{centroid:ids} & Shift query toward mean of examples & \texttt{q = $\alpha$\textperiodcentered q + (1-$\alpha$)\textperiodcentered mean(E[ids]); q /= $\|$q$\|$} & 1 mean & \cite{rocchio1971} \\
\texttt{from:A to:B} & Blend directional similarity & \texttt{scores = 0.5\textperiodcentered sim + 0.5\textperiodcentered (M @ (embed(B)-embed(A)))} & 1 matmul & \cite{mikolov2013} \\
\texttt{diverse} & MMR iterative selection ($\lambda$=0.7) & \texttt{score = $\lambda$\textperiodcentered rel - (1-$\lambda$)\textperiodcentered max\_sim} & $k \times n$ pairwise & \cite{carbonell1998} \\
\bottomrule
\end{tabular}
}% end resizebox
\caption{Five composable modulations and their costs. Costs are additional to the base similarity matmul.}
\label{tab:modulations}
\end{table}

One operation modifies the query vector before scoring: \texttt{centroid} shifts the query toward example embeddings. The remaining four operate on scores after the dot product.

\textbf{\texttt{suppress:}} subtracts directional similarity toward a named concept. When a corpus contains a large cluster of documentation and a smaller cluster of implementation, both may match a query like ``how the system works,'' but the documentation cluster can dominate the ranking. Suppression penalizes similarity to that direction, allowing the buried cluster to surface. Multiple \texttt{suppress:} tokens compose additively. Because suppression subtracts rather than projects orthogonally, it behaves like a soft negative query expansion, and when the suppressed direction is correlated with the query itself, relevant signal may also be reduced. Section~\ref{sec:suppress} evaluates this behavior.

\textbf{\texttt{decay:}} weights scores by recency using a configurable half-life. This is not a date filter; older content can still surface if sufficiently relevant. Short half-lives emphasize recent activity; longer half-lives preserve historical context.

\textbf{\texttt{centroid:}} shifts the query toward known-good examples. When a text query does not capture the desired facet, the agent can supply chunk IDs from prior results. The query vector shifts toward their mean, anchoring retrieval to a region of embedding space that words alone may not specify.

\textbf{\texttt{from:/to:} (trajectory)} blends a directional similarity component into scores. This applies embedding arithmetic to retrieval: the vector difference \texttt{embed(B) - embed(A)} defines a direction in embedding space, and candidates aligned with that direction score higher. \texttt{from:prototype to:production} surfaces content along a hardening trajectory without requiring those words in the query.

\textbf{\texttt{diverse}} changes selection rather than scoring. After all other modulations have reshaped scores, MMR iteratively selects from an oversample pool, penalizing each successive result for similarity to already-selected results. The effect is breadth rather than ten variations of the strongest signal.

Modulations execute in a fixed order regardless of token position: centroid (query shift), base similarity, trajectory, decay, suppress, diverse. Total cost is dominated by the initial matrix multiply.

The structural operators introduced in Section~\ref{sec:arch} (per-query clustering, centrality) compose with modulations: they operate on the modulated candidate set and expose computed columns for Phase~3. These are not evaluated in this paper.

\subsection{Interface}
\label{sec:interface}

\flexvec{} exposes modulations at three levels of abstraction: raw numpy operations, a token grammar, and SQL composition. Each level is usable independently, and together they form a composable interface.

\subsubsection{Python API}

\begin{lstlisting}[style=python]
scores = candidate_matrix @ query_vec
scores -= 0.5 * (candidate_matrix @ embed("database storage"))
scores *= 1.0 / (1.0 + days_ago / 7.0)
\end{lstlisting}

The modulations operate on any L2-normalized embedding matrix and do not depend on the SQLite integration or the materializer. The production API (\texttt{VectorCache.search()}) wraps these operations with token parsing and MMR selection.

\subsubsection{Token Grammar}

\flexvec{} provides a token grammar that maps to modulations with default parameters. A \texttt{pool:N} token controls the candidate pool size (default 500). Tokens are whitespace-delimited; prefix tokens (e.g., \texttt{suppress:}, \texttt{decay:}) map to modulations via a deterministic parser. Tokens can be written in any order; they execute in the fixed order described in Section~\ref{sec:modulations}. Tokens are shorthand with chosen defaults; the Python API (Section~3.4.1) allows full parameter control.

\begin{lstlisting}
diverse suppress:database sqlite storage decay:7 pool:1000
\end{lstlisting}

\subsubsection{Hybrid Retrieval}

Keyword search and modulated vector search compose via SQL JOIN. Chunks must match both the literal term and the semantic neighborhood:

\begin{lstlisting}[style=sql]
SELECT k.id, k.rank, v.score, m.content
FROM keyword('provenance') k
JOIN vec_ops('similar:file identity tracking diverse') v
    ON k.id = v.id
JOIN messages m ON k.id = m.id
ORDER BY v.score DESC LIMIT 10
\end{lstlisting}

\texttt{keyword()} uses FTS5 BM25 scoring via the same materializer pattern. The JOIN intersects both result sets. The agent controls the scoring and ranking logic via standard SQL.

\section{Evaluation}

All measurements are on 128-dimension embeddings (Nomic Embed v1.5~\cite{nomicembed2024}, Matryoshka truncation~\cite{matryoshka2022}). Desktop: Intel i9-13900KF (24 cores), 32GB RAM. numpy linked against OpenBLAS (default threading). All timings are warm-cache unless otherwise noted.

\subsection{Latency}

Latency breaks down across phases as shown in Table~\ref{tab:latency}, measured on the full 240,000-chunk corpus.

\begin{table}[H]
\centering
\begin{tabular}{lrl}
\toprule
Operation & Time & Scope \\
\midrule
Matrix multiply (240K $\times$ 128) & 5ms & Phase 2 only \\
Scoring + 3 modulations + MMR & 12ms & Phase 2 only \\
Full pipeline (pre-filter + scoring + compose) & 19ms & All phases \\
FTS5 keyword search & 18ms & --- \\
Hybrid (keyword JOIN vec\_ops) & 63ms & All phases \\
\bottomrule
\end{tabular}
\caption{Latency breakdown for retrieval operations on 240,000 chunks (128-dim, Intel i9-13900KF, 32GB RAM, numpy/OpenBLAS, warm cache). ``Phase~2 only'' measures numpy scoring without SQL pre-filter or compose. ``All phases'' includes SQL pre-filter, temp table write, and outer query.}
\label{tab:latency}
\end{table}

\subsection{Pre-filtering}

SQL pre-filtering reduces Phase~2 latency by narrowing candidates before scoring. Table~\ref{tab:prefil} shows the effect across five filter configurations, all with three composed modulations and MMR diversity selection.

\begin{table}[H]
\centering
\begin{tabular}{lrr}
\toprule
Pre-filter & Candidates & 3 mods + MMR \\
\midrule
Full corpus & 240,000 & 12ms \\
Non-tool, last 30 days & 61,000 & 12ms \\
Non-tool, last 7 days & 11,000 & 0.6ms \\
Non-tool, last 24 hours & 3,000 & 0.4ms \\
One project, 30 days & 37,000 & 5ms \\
\bottomrule
\end{tabular}
\caption{Effect of SQL pre-filtering on Phase~2 latency. Candidates are produced by a Phase~1 SQL pre-filter; the timings shown measure Phase~2 only (numpy scoring + three modulations + MMR) on those candidate sets. The full corpus and 61K rows show identical timing because the matmul dominates at these sizes. The latency drop becomes visible below $\sim$50K candidates.}
\label{tab:prefil}
\end{table}

\subsection{Scaling}

We constructed larger corpora by combining embeddings from multiple production datasets. Memory scales linearly with corpus size. Latency is sublinear at smaller sizes due to cache effects and becomes linear above $\sim$500K chunks. Table~\ref{tab:scaling} shows latency and memory as the corpus grows from 250K (the 240K production corpus combined with another dataset) to over one million chunks.

\begin{table}[H]
\centering
\begin{tabular}{lrrr}
\toprule
Corpus size & Base matmul & Full pipeline & Memory \\
\midrule
250K & 5ms & 19ms & 122MB \\
500K & 7ms & 37ms & 244MB \\
750K & 15ms & 73ms & 366MB \\
1M & 17ms & 82ms & 514MB \\
\bottomrule
\end{tabular}
\caption{End-to-end latency and memory scaling from 250K to 1M chunks (warm cache). ``Base matmul'' is the Phase~2 matrix multiply component; ``Full pipeline'' includes SQL pre-filter, Phase~2 scoring + three modulations + MMR, temp table write, and outer SQL compose. Memory is the embedding matrix only.}
\label{tab:scaling}
\end{table}

\subsection{Behavioral Validation}
\label{sec:validation}

All modulations use their token grammar defaults: \texttt{suppress:} weight $w=0.5$, \texttt{centroid:} blend $\alpha=0.5$, \texttt{from:/to:} blend 0.5/0.5, \texttt{decay} default half-life 30 days (configurable via \texttt{decay:N}), \texttt{diverse} $\lambda=0.7$ with 3x oversample pool, temp table size $K=500$. Scores are not renormalized between modulations.

Three metrics are used to measure behavioral change:

\begin{itemize}[leftmargin=*]
\item \textbf{RBO (Rank-Biased Overlap)}~\cite{rbo2010}: similarity between two ranked lists, weighted toward the top. RBO=1.0 means identical rankings; low RBO indicates the modulation produced a substantially different result set.
\item \textbf{ILS (Intra-List Similarity)}: mean pairwise cosine similarity among the top-$K$ results. Lower ILS indicates greater diversity.
\item \textbf{Centroid similarity}: cosine similarity between each result and the centroid of the seed examples. Higher values indicate the query expansion shifted results toward the intended facet.
\end{itemize}

We assess correctness at two levels: algebraic (scores match formulas) and behavioral (rankings change as intended). For behavioral validation, we tested against four BEIR datasets: SciFact (5,183 documents), NFCorpus (3,633), SCIDOCS (25,657), and FiQA (57,638). Baseline nDCG@10 ranged from 0.13 (NFCorpus) to 0.60 (SciFact) across the four corpora. The low baseline on NFCorpus likely reflects the combination of short queries against a specialized biomedical corpus at reduced dimensionality; modulation behaviors were present across all four corpora regardless of baseline strength.

\begin{table}[H]
\centering
\begin{tabular}{lll}
\toprule
Modulation & Effect & nDCG@10 retention \\
\midrule
\texttt{diverse} & ILS reduction 10--40\% across all corpora & See Table~\ref{tab:diverse} \\
\texttt{suppress:X} & RBO vs baseline 0.19--0.41 & --- \\
\texttt{decay:7} & Results 42--50 days more recent on average & --- \\
\texttt{centroid:ids} & Centroid similarity +0.05 to +0.12 & --- \\
\texttt{from:A to:B} & RBO vs baseline 0.08--0.25 & --- \\
\bottomrule
\end{tabular}
\caption{Behavioral validation on four BEIR datasets (30 queries per dataset). Modulations produce measurable ranking changes. nDCG retention is reported only for \texttt{diverse} (Table~\ref{tab:diverse}); for the other modulations, which intentionally alter the retrieval objective, we emphasize behavioral diagnostics rather than task-level effectiveness claims.}
\label{tab:behav}
\end{table}

\begin{table}[H]
\centering
\begin{tabular}{lrrrr}
\toprule
Corpus & Baseline nDCG@10 & \texttt{diverse} nDCG@10 & Retention & ILS reduction \\
\midrule
SciFact & 0.60 & 0.56 & 93\% & 12\% \\
FiQA & 0.41 & 0.33 & 80\% & 28\% \\
SCIDOCS & 0.18 & 0.15 & 83\% & 18\% \\
NFCorpus & 0.13 & 0.08 & 59\% & 40\% \\
\bottomrule
\end{tabular}
\caption{Per-corpus \texttt{diverse} nDCG@10 retention and ILS reduction. Tightly clustered corpora (NFCorpus) pay a higher diversity cost; broadly distributed corpora (SciFact) retain more relevance. The 59\% retention on NFCorpus reflects the density of its biomedical domain; MMR's diversity penalty is proportionally larger when candidates are tightly clustered.}
\label{tab:diverse}
\end{table}

The test evaluates 30 queries per dataset as a lightweight behavioral validation suite. A separate algebraic test suite verifies that each modulation produces scores matching its formula to floating-point precision: across four corpora, 1,840 individual score comparisons produced zero mismatches. Together, the behavioral and algebraic tests demonstrate that the operations are correct and produce their intended effects; they are not a competitive retrieval benchmark. Reproduction details are in Appendix~A.

\section{Case Studies}

\subsection{Suppressing Dominant Clusters}
\label{sec:suppress}

AI coding agents accumulate session history (prompts, responses, tool calls, file edits) in corpora that grow daily. When an agent searches this history, embedding similarity alone can return results from the dominant cluster rather than the intended one. Content that \textit{describes} a system and content that \textit{implements} it use the same vocabulary and occupy the same region of embedding space. The descriptive cluster is typically larger: documentation rewrites generate many variations, landing page iterations produce dozens of drafts, and each contains the same key terms as the actual architecture. Baseline cosine similarity returns the highest-scoring content, which tends to come from the larger cluster.

Metadata filters are limited when the distinction is semantic rather than structural. Both clusters are assistant messages, span multiple projects, and share the same time period. The difference is what the content is \textit{for}, not where it lives in the schema.

Suppression addresses this by subtracting directional similarity toward named topics, reshaping the score surface so that the buried cluster surfaces.

An agent queried the production corpus for ``how the system works.'' The baseline returned five results from the descriptive cluster: page layouts, tagline drafts, documentation structure, positioning discussions. The actual architecture (identity layers, server operations, production debugging) scored lower because the descriptive cluster was larger and used the same vocabulary. Adding two \texttt{suppress:} tokens to the same query reshapes the score surface:

\begin{lstlisting}[style=sql]
SELECT v.id, v.score, m.content
FROM vec_ops(
    'similar:how the system works architecture
     diverse
     suppress:website landing page design tagline
     suppress:documentation readme community post',
    'SELECT id FROM messages
     WHERE type = ''assistant'' AND length(content) > 300') v
JOIN messages m ON v.id = m.id
ORDER BY v.score DESC LIMIT 5
\end{lstlisting}

Two \texttt{suppress:} tokens target the descriptive cluster, one for marketing copy, one for documentation. Each embeds its text and subtracts that directional similarity from scores. The multi-word syntax captures a semantic direction, not individual keywords.

\FloatBarrier
\begin{table}[H]
\centering
\small
\begin{tabular}{cp{5.8cm}p{5.8cm}}
\toprule
 & Baseline & Suppressed \\
\midrule
1 & UI style iteration \hfill\textit{0.70} & Identity layer architecture \hfill\textit{0.18} \\[4pt]
2 & Marketing tagline draft \hfill\textit{0.67} & Server lifecycle debugging \hfill\textit{0.14} \\[4pt]
3 & Documentation site structure \hfill\textit{0.65} & Background worker failure analysis \hfill\textit{0.07} \\[4pt]
4 & Product positioning discussion \hfill\textit{0.65} & Rendering pipeline implementation \hfill\textit{0.05} \\[4pt]
5 & Community post draft \hfill\textit{0.65} & Platform detection and branching logic \hfill\textit{0.03} \\
\bottomrule
\end{tabular}
\caption{Top-5 results for ``how the system works'' with and without suppression of descriptive content. Baseline scores cluster at 0.65--0.70. Suppressed scores spread from 0.03 to 0.18. None of the suppressed results appeared in the baseline top results. Scores are not comparable across configurations because suppression changes the scoring function.}
\label{tab:suppress}
\end{table}

The query intent is ``how the system works'': the caller wants to understand the system's architecture and operation. The baseline results all describe how the system is \textit{presented} (UI styling, taglines, documentation structure, positioning) because the descriptive cluster is larger and uses the same vocabulary as the query. The suppressed results surface what the baseline buried: the identity layer's data model, server lifecycle behavior, background processing architecture, and deployment-time branching logic.
\FloatBarrier
\subsection{Pre-filtering and Scaling}

An agent searched for ``file identity tracking'' restricted to a message type comprising approximately 1\% of the corpus (3,961 of 240,000 chunks). With a post-filter on this type, the query returned 1 result. The vector search selected 200 candidates by cosine similarity first, then the post-filter kept only those that matched the target type. Only 1 of the 200 top-cosine candidates matched. The other 3,960 never entered scoring.

This is pool starvation, a known limitation of post-filtered retrieval~\cite{filteredann2025} that becomes acute when the target subset is sparse relative to the candidate pool. The failure occurs silently: the system returns results, but they do not correspond to the target subset.

Phase~1 fixes this. A SQL pre-filter scopes candidates before scoring:

\begin{lstlisting}[style=sql]
SELECT id FROM messages WHERE type = 'target_type'
\end{lstlisting}

All 3,961 matching chunks enter the embedding matrix. Cosine similarity and modulations operate on the correct candidate set.

The same mechanism reduces scoring cost. An agent searching for recent architecture decisions does not score 240,000 chunks. It writes:

\begin{lstlisting}[style=sql]
SELECT id FROM messages
WHERE type = 'assistant' AND project = 'core'
  AND created_at > date('now', '-7 days')
\end{lstlisting}

This reduces the candidate set to a few thousand before any vector math occurs. Three composed modulations then execute in under 1ms on the narrowed set (Table~\ref{tab:prefil}). The effectiveness of this approach depends on the pre-filter's selectivity. The architecture is agnostic to how the candidate set is produced; it requires only that a set of chunk IDs enters Phase~2.

Pre-filtering decides which candidates enter scoring. Modulation reshapes scores among those candidates.

\FloatBarrier
\section{Related Work}
\label{sec:relwork}

\flexvec{} takes Rocchio relevance feedback~\cite{rocchio1971}, word embedding arithmetic~\cite{mikolov2013}, and MMR diversity~\cite{carbonell1998} and exposes them as composable primitives inside a single SQL statement. This section positions the architecture relative to systems that compose similar operations differently.

PyTerrier~\cite{pyterrier2020} composes retrieval transformers --- retrievers, rerankers, and fusion operators --- using algebraic pipeline expressions built from overloaded Python operators. The algebra operates across pipeline stages; each stage is a self-contained transformer with its own scoring logic.

Qdrant's Query Points API~\cite{qdrant2024} composes dense search, sparse search, recommendation, discovery, metadata filters, and score-boosting formulas through a nested prefetch pipeline with fusion and reranking. Vespa~\cite{vespa2024} composes per-document scoring functions within rank profiles defined at deploy time; query-time tensors can parameterize these expressions, and its global-phase ranking supports cross-hit operations on the merged result set. Both systems compose predefined query modes rather than exposing the score array for caller-side arithmetic.

sqlite-vec~\cite{sqlitevec2024} provides brute-force KNN inside SQLite via C virtual tables, with filtering scoped to metadata columns and partition keys declared at table creation and a limited set of comparison operators. pgvector~\cite{pgvector2024} extends PostgreSQL with HNSW and IVFFlat indexes, allowing standard SQL WHERE clauses alongside vector distance operators; the SQL predicates and vector scoring share one query planner, and highly selective filters can force fallback from the ANN index to brute-force scan. MyScale~\cite{myscale2024} integrates vector indexing into a ClickHouse-forked engine with full SQL support, using columnar pre-filtering before ANN traversal. In all three systems, candidate selection and vector scoring are evaluated within a single execution process.

RAGdb~\cite{ragdb2026} consolidates hybrid retrieval into a single SQLite container using TF-IDF with substring boosting. Faiss~\cite{faiss2024} provides GPU-accelerated similarity search with quantization and partitioning; at the corpus sizes described here, the additional index infrastructure is not required.

DEO~\cite{deo2026} decomposes queries into positive and negative components and optimizes the query embedding at inference time via contrastive loss, modifying the query vector itself rather than the score array. IMRNNs~\cite{imrnns2026} uses the phrase ``embedding modulation'' for learned MLP adapters that dynamically adjust query and document embeddings at inference time. Both operate in embedding space with learned or optimized parameters; PEM operates post-matmul on the score array, without learned parameters and without an optimization step.

Recent RAG pipelines frequently incorporate LLM-based hypothetical document generation (e.g., HyDE~\cite{hyde2023}), late interaction models (ColBERT~\cite{colbert2020}), or neural rerankers applied after initial retrieval. These operate at higher layers of the retrieval stack than the score-level modulations described here.

In the systems surveyed, candidate selection and vector scoring are either co-located within one query planner or filtering is constrained to columns declared at index creation. \flexvec{} separates these concerns: an arbitrary SQL query materializes the candidate set before it enters the scoring engine. The scoring engine then exposes the score array for caller-composed modulations before selection.

\section{Limitations}
\label{sec:limitations}

\begin{itemize}[leftmargin=*]
\item \textbf{Scope.} \flexvec{} is not presented here as a replacement for ANN-based systems at web scale; it requires the candidate embedding matrix in memory for direct access. SQL pre-filtering narrows the candidate set before scoring. Corpora without structured metadata score the full matrix.
\item \textbf{Retrieval quality.} This paper evaluates latency, architecture, algebraic correctness, and behavioral effects of individual modulations (Section~\ref{sec:validation}). Comprehensive retrieval quality evaluation on production corpora, including comparison to reranking baselines, is future work.
\item \textbf{Scale and dimensionality.} All benchmarks use 128-dimension embeddings. At 1M chunks, three composed modulations execute in 82ms. Higher-dimension models will increase matmul cost. The system was previously deployed with MiniLM at 384 dimensions; migration to Nomic Embed at 128-dimensional Matryoshka truncation produced no informally observed degradation in production use. Formal comparison across models and dimensionalities is future work.
\item \textbf{Pre-filter depends on structured metadata.} The SQL pre-filter is effective because the production corpus has typed columns (message type, project, timestamps). Corpora without structured metadata fall back to scoring the full corpus.
\item \textbf{Materializer robustness.} The SQL materializer uses query rewriting rather than a full SQL parser. A native SQLite extension would cede control of execution order to SQLite's query planner; the materializer preserves the explicit Phase~1$\rightarrow$2$\rightarrow$3 pipeline at the cost of regex-based query rewriting. In an agent-facing setting, the failure mode is explicit rewrite failure rather than silent misexecution: the system returns an error, the agent rewrites the query, and retries. Across approximately 6,500 agent queries over six weeks, no materializer rewrite failures were observed in normal use, though edge cases with deeply nested SQL may still exist.
\item \textbf{Agent error rate.} Agent SQL composition errors decreased from approximately 10\% to 4\% after improvements to schema documentation and error guidance. The most common errors are incorrect column references, malformed pre-filter subqueries that return no rows, and occasional table alias conflicts in complex JOINs. These are SQL composition errors by the agent, not failures in the modulation or materializer. The system returns an error via MCP, and the agent typically self-corrects on retry.
\item \textbf{Data provenance.} The production corpus is a single-user AI coding session history. The BEIR datasets used for behavioral validation are public benchmarks.
\end{itemize}

\section{Conclusion}

\flexvec{} exposes the embedding matrix and score array as a programmable surface, allowing arithmetic operations on both before selection. The three-phase architecture (SQL pre-filter, numpy modulation, SQL composition) integrates this into standard SQL via a query materializer: the agent's statement describes what to retrieve, and the materializer decides how to execute it. The result is that retrieval can be expressed as a composed query rather than a single search call.

The operator set is extensible, and new query-time operations can be exposed through the same interface pattern. \flexvec{} achieves \mbox{sub-100ms} latency at corpora up to one million chunks on a desktop CPU. Formal retrieval quality evaluation across diverse corpora and analysis of agent query patterns are future work. \flexvec{} is available at \url{https://github.com/damiandelmas/flexvec} (MIT) and is the retrieval kernel of \flex{}, a local search and retrieval system for AI agents.

\bibliographystyle{plain}

\begin{thebibliography}{19}

\bibitem{rocchio1971}
J.J.\ Rocchio.
Relevance Feedback in Information Retrieval.
\textit{The SMART Retrieval System}, 1971.

\bibitem{mikolov2013}
T.\ Mikolov, K.\ Chen, G.\ Corrado, and J.\ Dean.
Efficient Estimation of Word Representations in Vector Space.
\textit{ICLR}, 2013.

\bibitem{carbonell1998}
J.\ Carbonell and J.\ Goldstein.
The Use of MMR, Diversity-Based Reranking for Reordering Documents and Producing Summaries.
\textit{SIGIR}, 1998.

\bibitem{pyterrier2020}
C.\ Macdonald and N.\ Tonellotto.
Declarative Experimentation in Information Retrieval using PyTerrier.
\textit{ICTIR}, 2020.

\bibitem{qdrant2024}
Qdrant.
Hybrid Queries.
\url{https://qdrant.tech/documentation/concepts/hybrid-queries/}, 2024.

\bibitem{sqlitevec2024}
A.\ Garcia.
sqlite-vec: A vector search SQLite extension.
\url{https://alexgarcia.xyz/sqlite-vec/}, 2024.

\bibitem{ragdb2026}
A.B.\ Khalid.
RAGdb: A Zero-Dependency, Embeddable Architecture for Multimodal Retrieval-Augmented Generation on the Edge.
\textit{arXiv:2602.22217}, 2026.

\bibitem{filteredann2025}
Y.\ Lin et al.
Survey of Filtered Approximate Nearest Neighbor Search over the Vector-Scalar Hybrid Data.
\textit{arXiv:2505.06501}, 2025.

\bibitem{deo2026}
T.\ Lee, J.\ Park, S.\ Hwang, and J.\ Jang.
DEO: Training-Free Direct Embedding Optimization for Negation-Aware Retrieval.
\textit{arXiv:2603.09185}, 2026.

\bibitem{hyde2023}
L.\ Gao et al.
Precise Zero-Shot Dense Retrieval without Relevance Labels.
\textit{ACL}, 2023.

\bibitem{colbert2020}
O.\ Khattab and M.\ Zaharia.
ColBERT: Efficient and Effective Passage Search via Contextualized Late Interaction over BERT.
\textit{SIGIR}, 2020.

\bibitem{rbo2010}
W.\ Webber, A.\ Moffat, and J.\ Zobel.
A Similarity Measure for Indefinite Rankings.
\textit{ACM TOIS}, 2010.

\bibitem{nomicembed2024}
Z.\ Nussbaum et al.
Nomic Embed: Training a Reproducible Long Context Text Embedder.
\textit{arXiv:2402.01613}, 2024.

\bibitem{matryoshka2022}
A.\ Kusupati et al.
Matryoshka Representation Learning.
\textit{NeurIPS}, 2022.

\bibitem{faiss2024}
M.\ Douze et al.
The Faiss Library.
\textit{arXiv:2401.08281}, 2024.

\bibitem{vespa2024}
Vespa Team.
Vespa: Tensor Ranking Expressions.
\url{https://docs.vespa.ai/en/ranking/ranking-intro.html}, 2024.

\bibitem{imrnns2026}
Y.\ Saxena, A.\ Padia, K.\ Gunaratna, and M.\ Gaur.
IMRNNs: An Efficient Method for Interpretable Dense Retrieval via Embedding Modulation.
\textit{arXiv:2601.20084}, 2026.

\bibitem{pgvector2024}
A.\ Kane.
pgvector: Open-source vector similarity search for Postgres.
\url{https://github.com/pgvector/pgvector}, 2024.

\bibitem{myscale2024}
MyScale.
MyScaleDB: An open-source, high-performance SQL vector database built on ClickHouse.
\url{https://github.com/myscale/myscaledb}, 2024.

\end{thebibliography}

\newpage
\appendix

\section{Reproduction}
\label{app:repro}

The behavioral validation suite runs 52 property tests across four BEIR datasets: SciFact (5,183 documents), NFCorpus (3,633), SCIDOCS (25,657), and FiQA (57,638).

\textbf{Setup:}

\begin{lstlisting}
python tests/build_beir_db.py          # downloads all 4, embeds (~1 hour CPU)
pytest tests/test_tokens_beir.py       # 52 tests, ~20 seconds
\end{lstlisting}

\textbf{Dependencies:}

\begin{tabular}{lll}
\toprule
Package & Version & Role \\
\midrule
numpy & $\geq$1.24 & Matrix operations, scoring \\
onnxruntime & $\geq$1.17 & Embedding inference (CPU) \\
tokenizers & $\geq$0.15 & Nomic tokenizer \\
\bottomrule
\end{tabular}

\textbf{Embedding model:} \texttt{nomic-embed-text-v1.5} at 128-dim Matryoshka truncation (ONNX int8 export). BEIR datasets are downloaded directly from the UKP mirror (no HuggingFace dependency). Documents and queries must use the same embedder to ensure consistent embedding space.

\textbf{Version pinning:} For exact reproduction, record Python and dependency versions (e.g., \texttt{python -V} and \texttt{pip freeze}) in the environment used to run the suite.

\textbf{What the tests verify:} Each modulation token is tested independently across four corpora spanning science, biomedical, financial, and computer science domains. Tests assert that modulations produce measurable ranking changes (measured by RBO divergence from baseline). For \texttt{diverse}, tests additionally assert nDCG@10 retention within specified thresholds. For the other modulations, which intentionally alter the retrieval objective, the tests verify behavioral change rather than task-level retention. The test evaluates 30 queries per dataset (by insertion order, not random selection).

\textbf{Algebraic correctness:} A separate test suite (\texttt{test\_algebraic.py}) verifies that each modulation produces scores matching its formula exactly. For each of five operations (suppress, multi-suppress, trajectory, decay, centroid), the test computes expected scores via the formula and compares against actual query output. Across four corpora: 1,840 score comparisons, zero mismatches beyond floating-point tolerance (1e-3).

\textbf{Caveats:} Timestamps in the BEIR databases are synthetic (90-day uniform spread), so \texttt{decay:N} tests verify the mechanism but not real temporal ranking. The 30-query subset runs slightly optimistic relative to the full query set (e.g., SciFact nDCG@10 is 0.60 on 30 queries vs 0.55 on all 300).

Source code: \url{https://github.com/damiandelmas/flexvec}

\section{MCP Server Instructions}
\label{app:mcp}

The following is an excerpt of the instructions injected into the AI agent's context when it connects to the MCP server.

\begin{lstlisting}
Flex offers a single endpoint for all operations: flex_search.
Two parameters: query (SQL or @preset) and cell (cell name).

The retrieval pipeline for each query is SQL -> vec_ops -> SQL.
Phase one narrows the corpus with SQL. Phase two scores it with
embeddings. Phase three composes the final result with SQL.

PHASE 1: SQL PRE-FILTER

Phase 1 narrows the candidate set before any vector operations
occur. Push known constraints here, not into a WHERE clause
after vector operations.

    SELECT id FROM messages WHERE type = 'user_prompt'
    SELECT id FROM messages WHERE session_id LIKE 'abc123%'

PHASE 2: VECTOR OPERATIONS (vec_ops)

Phase 2 scores the candidate set from Phase 1 using embeddings.
Tokens let you reshape the scoring before selection. Tokens are
independent operations that compose freely and stack into one
pass.

    vec_ops('similar:query_text tokens', 'pre_filter_sql')

vec_ops appears in FROM position like a table source, but is a
pseudo-function recognized by the materializer before SQLite
executes the query. Always use after FROM or JOIN:

    FROM vec_ops('similar:query') v

Modulation Tokens

Tokens can be written in any order:
    'similar:auth diverse suppress:jwt decay:7'
They execute in a fixed order (centroid, similarity, trajectory,
decay, suppress, diverse) regardless of appearance in the string.

  diverse       MMR -- penalizes similarity to already-selected.
  suppress:TEXT  Contrastive search. Demotes chunks similar to TEXT.
  decay:N       Temporal decay with N-day half-life.
  centroid:ids  Centroid search from example chunk IDs.
  from:A to:B   Trajectory -- direction vector between concepts.
  pool:N        Candidate pool size (default 500).

RECIPES

Exact term (FTS5 -- domain name, filename, error, UUID):

    SELECT k.id, k.rank, k.snippet, c.content
    FROM keyword('term') k
    JOIN chunks c ON k.id = c.id
    ORDER BY k.rank DESC LIMIT 10
    -- keyword() is a table source -- always use after FROM or JOIN.
    -- Returns (id, rank, snippet). rank is positive -- higher = better.
    -- Special chars (dots, operators) handled automatically via
    -- fallback quoting.

Hybrid intersection (only chunks matching BOTH keyword and
semantic):

    SELECT k.id, k.rank, v.score, c.content
    FROM keyword('auth') k
    JOIN vec_ops('similar:authentication patterns') v
        ON k.id = v.id
    JOIN chunks c ON k.id = c.id
    ORDER BY v.score DESC
    LIMIT 10

PHASE 3: SQL COMPOSITION

Phase 3 joins scored results back to views for grouping,
boosting, filtering, and pagination.

    SELECT v.id, v.score, c.content
    FROM vec_ops('similar:authentication') v
    JOIN chunks c ON v.id = c.id
    ORDER BY v.score DESC LIMIT 10

METHODOLOGY

Start with @orient. Every cell describes itself.
Feel the data before writing complex queries.
Discover then narrow. Pivot when the mode shifts.
\end{lstlisting}

\section{@orient Preset SQL}
\label{app:orient-sql}

The \texttt{@orient} preset is a multi-query SQL script stored in the database. Each \texttt{-{}- @query:} section produces one key of the output. The preset uses \texttt{pragma\_table\_info()} to discover view columns at runtime, ensuring the output reflects the current schema without hardcoding.

\begin{lstlisting}[style=sql]
-- @query: now
SELECT datetime('now', 'localtime') as now,
       'UTC' || printf('%+d',
       cast((julianday('now','localtime')
       - julianday('now')) * 24 as integer)) as timezone;

-- @query: about
SELECT value as description FROM _meta
WHERE key = 'description';

-- @query: shape
SELECT 'chunks' as what, COUNT(*) as n FROM _raw_chunks
UNION ALL
SELECT 'sources', COUNT(*) FROM _raw_sources;

-- @query: query_surface
SELECT 'view' as kind, m.name as name,
    GROUP_CONCAT(p.name, ', ') as columns,
    CASE m.name
        WHEN 'chunks' THEN 'UNIFIED surface -- all chunks.
            type: user_prompt|assistant|tool_call|file.'
        WHEN 'messages' THEN 'Message chunks only.'
        WHEN 'sessions' THEN 'Sources with graph intelligence.'
        ELSE ''
    END as note
FROM sqlite_master m, pragma_table_info(m.name) p
WHERE m.type = 'view'
GROUP BY m.name
UNION ALL
SELECT 'table_function', 'vec_ops', 'id, score',
    'Semantic retrieval -- use after FROM/JOIN.'
UNION ALL
SELECT 'table_function', 'keyword', 'id, rank, snippet',
    'FTS5 keyword search.'
ORDER BY kind, name;

-- @query: presets
SELECT name, description, params FROM _presets ORDER BY name;
\end{lstlisting}

\section{@orient Output}
\label{app:orient-output}

The following is an excerpt of the \texttt{@orient} output on the production corpus ($\sim$240K chunks).

\begin{lstlisting}
now:        2026-03-12 16:24:00

about:      Agentic coding conversation history. Sessions, messages,
            tool calls, and output. Each source is a session,
            each chunk is a message.

shape:      chunks: ~240,000
            sources: ~4,000

query_surface:

  Views (primary query surface)

  - chunks -- id, content, timestamp, created_at, type,
    session_id, position, project, tool_name, file, section,
    ext, child_session_id, agent_type, file_uuids
    UNIFIED surface -- all chunks (messages + files).
    type: user_prompt|assistant|tool_call|file.

  - files -- id, content, timestamp, created_at, session_id,
    file, title, chunk_position, ext
    File body sub-chunks only.

  - messages -- id, content, timestamp, created_at, session_id,
    position, project, title, message_count, tool_name,
    target_file, success, cwd, type, child_session_id,
    agent_type, file_uuids, file_body
    Message chunks only (no file body sub-chunks).

  - sessions -- session_id, project, title, message_count,
    start_time, duration, started_at, ended_at, chunk_count,
    fingerprint_index, centrality, is_hub, is_bridge,
    community_id, community_label
    Sources with graph intelligence, fingerprints.

  Table Functions

  - keyword(term) -> (id, rank, snippet)
    FTS5 keyword search.

  - vec_ops('similar:QUERY TOKENS', pre_filter_sql)
        -> (id, score)
    Semantic retrieval.

presets:
  @digest           Multi-day activity summary (days=7)
  @file             Sessions that touched a file (path required)
  @file-search      BM25 over file bodies (query required)
  @genealogy        Concept lineage -- timeline, hubs, excerpts
  @orient           Full cell orientation
  @sprints          Work sprints detected by 6h gaps
  @story            Session narrative (session required)
\end{lstlisting}

\end{document}